\documentclass[aps,prb,preprint,groupedaddress,showpacs]{revtex4}
\usepackage{graphicx}
\begin{document}
\title{
Theory of Zero-Bias Anomaly in Magnetic Tunnel Junctions:
Inelastic Tunneling via Impurities
}
\author{L. Sheng$^{1}$, D. Y. Xing$^{1}$ and D. N. Sheng$^{2}$}
\address{$^1$National Laboratory of Solid State Microstructures, Nanjing University, Nanjing 210093, China\\
$^2$Department of Physics and Astronomy, California State University, Northridge, CA 91330, USA}
\begin{abstract}
Using the closed-time path integral approach, we
nonperturbatively study inelastic tunneling of electrons
via magnetic impurities in the barrier
accompanied by phonon emission in a magnetic tunnel
junction. The spectrum density of phonon emission
is found to show a power-law infrared singularity
$\sim\omega^{-(1-g)}$ with $g$ the dimensionless electron-phonon coupling.
As a consequence, the tunneling conductance $G(V)$ increases with bias
voltage $\vert V\vert$ as $G(V)-G(0)\sim\vert V\vert^{2g}$, exhibiting a
discontinuity in slope at $V=0$ for $g\le 0.5$.
This theory can reproduce both cusp-like and non-cusp-like
feature of the zero-bias anomaly of tunneling resistance and magnetoresistance
widely observed in experiments.
\end{abstract}
\mbox{}

\pacs{PACS Number: 72.25.Ba, 75.47.-m, 73.40.Rw,72.10.Di}
\maketitle The tunneling magnetoresistance (TMR) effect in a
magnetic junction composed of two ferromagnetic (FM) electrodes
separated by an insulating layer is presently of great interest
because of its potential of device applications such as random
access memories and magnetic
sensors~\cite{s1,s2,s3,s4,s5,s6,s7,s8,s9,add7,s10}. While the TMR
effect has been attributed to spin polarization of conduction
electrons in the FM electrodes~\cite{s1,s2,s3}, its novel voltage
dependence is still puzzling. The tunneling resistance and TMR
ratio generally decrease with increasing bias voltage, often
exhibiting a cusp-like feature at zero
bias~\cite{s1,s5,s6,s7,s8,s9,add7,s10}. This effect, usually
called the zero-bias anomaly (ZBA)~\cite{s6}, was not only widely
observed in transition-metal magnetic junctions, but also exists
in semiconductor GaMnAs/AlAs/GaMnAs magnetic
junctions~\cite{add3}.

It is found that the ZBA is  very sensitive to the insulating material
of the barrier. In earlier measurements~\cite{s1} on Fe/Ge/Co junctions,
the characteristic voltage $V_{1/2}$ that halves the TMR ratio is about
3 mV. It was reported recently that for FM/Al$_2$O$_3$/FM with Fe, Co,
Ni, or their alloys as the FM electrodes, $V_{1/2}$ is usually one hundred
to several hundred mV~\cite{s5,s6,s7,s8,s9}, and for Ni/NiO/Co junctions
$V_{1/2}$ is a few ten mV~\cite{s10}. Junctions made of the same FM
electrodes and identical but differently prepared insulator vary
considerably concerning the voltage dependence of the resistance and
TMR~\cite{s9}. It was reported that the ZBA in the semiconductor magnetic
tunnel junctions is much more striking with $V_{1/2}$ about $3$mV~\cite{add3}.
Interestingly, in a very recent experiment~\cite{s11} of the TMR between a
Co sample and a CoFeSiB tip through a vacuum barrier, the ZBA did not occur
for bias voltage upto a volt.

Understanding of the ZBA is not only
important for optimizing the TMR in practical applications, but also
provides critical insight into the underlying physics of spin-dependent
transport. There exist several possible explanations for the
anomaly. The simplest one may involve the energy dependence of the
electron density of states (DOS) and elastic transmission matrix
elements~\cite{s3,s14,sSheng}. Unfortunately, by comparison with
experimental data, the TMR ratio calculated from such a model decreases
too slowly at low bias~\cite{s7}. The second mechanism is that the hot electrons
from the emitting electrode may be scattered by local magnetic moments
at the interfaces through $s$-$d$ interaction~\cite{s6}, which
contributes to the conductance a term $\Delta G\sim\vert V\vert$ at low
bias voltage $V$. This theory succeeded in explaining some experimental
data. The third mechanism might be based upon the electron tunneling
assisted by magnons and phonons~\cite{s7,add1,add2,add4}, which was found
to contribute $\Delta G\sim\vert V\vert^{3/2}$ for FM magnons,
$\vert V\vert^{2}$ for surface antiferromagnetic magnons,
and $\vert V\vert^4$ for phonons. This mechanism may
explain some experimental data without the cusp-like feature
at zero bias~\cite{add4},
but has the difficulty to interpret the cusp-like feature of the ZBA
observed in many other experiments~\cite{s5,s6,s7,add7,add3}.
On contrary to these intrinsic mechanisms,
the effect of  extrinsic nonmagnetic and magnetic impurities
was emphasized in many works~\cite{s8,add7,s10,s14,add4,add5,add6,add8,add9,add10}.
The voltage dependence caused by the elastic electron tunneling via impurities
was studied theoretically in a few works~\cite{s10,s14,add8}. However, none of them
reproduced the essential features of the ZBA.
Based upon the results of an earlier theory~\cite{add11},
some authors~\cite{add7,add10} argued that, in the presence of
inelastic scattering, while single-impurity tunneling processes
are still elastic, electron tunneling via multiple-impurity chains
may contribute an inelastic
power-law conductance $\Delta G\sim\vert V\vert^{p}$
with $p=4/3$  $(5/2)$ for two-impurity (three-impurity) processes.
Presently, the relevant mechanism for the ZBA is still under debate.

In this work, a combined effect of the magnetic impurities in the
barrier and the electron-phonon interaction is proposed as a
possible mechanism for the ZBA. What we consider in this model is
not a simple addition of the elastic tunneling of electrons via
the impurities and the inelastic one assisted by
phonons~\cite{add4}, but a novel inelastic electron tunneling due
to the strong interplay between them, in which the electron
tunneling via impurities is always accompanied by phonon emission.
It is found that the spectrum density of phonon emission has a
power-law infrared singularity $\sim\omega^{-(1-g)}$ with $g$ the
dimensionless electron-phonon coupling. As a consequence, the
conductance via impurities $G^{\mbox{\tiny I}}(V)$ obeys a
power-law voltage dependence $G^{\mbox{\tiny I}}(V)\sim\vert
V\vert^{2g}$ with a discontinuity in slope at $V=0$ for $g\le
0.5$. The conductance exponent $2g$ may be either smaller or
greater than unity, being different from the earlier
theories~\cite{s6,add1,add2,add4,add11} where the conductance
exponents are never smaller than unity. The calculated results can
reproduce both cusp-like and non-cusp-like features of ZBA
observed in experiments. An experiment based upon the isotope
effect is suggested to further verify this theory.

The model Hamiltonian of a magnetic tunnel junction is written as
$H=H_{\mbox{\tiny L}}+H_{\mbox{\tiny R}}+H_{\mbox{\tiny C}}+H_{\mbox{\tiny T}}$.
Here, $H_{\mbox{\tiny L(R)}}=\sum_{ks}\epsilon_{ks}^{\mbox{\tiny L(R)}}
a_{ks}^{\mbox{\tiny L(R)}\dagger} a^{\mbox{\tiny L(R)}}_{ks}$ is the
Hamiltonian of the electrons in the left (right) electrodes.
The third term is the Hamiltonian of the electrons on the magnetic impurities,
\begin{eqnarray}
H_{\mbox{\tiny C}}&=&\sum\limits_{lss'}(E_0\delta_{ss'}
-J\mbox{\boldmath{$\sigma$}}_{ss'}\cdot{\bf S}_l) d_{ls}^{\dagger} d_{ls'}
+\sum\limits_{q\lambda}\omega_{q\lambda}b_{q\lambda}^{\dagger} b_{q\lambda}
+\sum\limits_{q\lambda ls}M_{q\lambda}e^{i{\bf q}\cdot{\bf R}_l}
\left(b_{q\lambda} + b_{-q\lambda}^\dagger\right)d_{ls}^\dagger d_{ls}\ .
\label{HC}
\end{eqnarray}
Here, $d_{ls}$ annihilates an electron with spin $s$ on magnetic impurity $l$, and
$b_{q\lambda}$ annihilates an acoustic phonon of wave vector $q$
and polarization $\lambda$ in the system.
The magnetic impurities in the barrier are presumably the atoms of
FM metals of the electrodes mixed into the barrier.
Some of their outer-shell electrons are itinerant, and the others form localized spins.
$J$ represents the $s$-$d$ like interaction
between the itinerant electrons and localized spin ${\bf S}_l$
with $\hat{\mbox{\boldmath{$\sigma$}}}$ the Pauli matrix.
The spins of the transition-metal ferromagnets under consideration are
usually greater than $1/2$, and can be treated as classical approximately.
Since the impurities are also a part of the
lattice, electrons scattering with the impurities will generally
cause emission or absorption of phonons of the system, which is described
by the electron-phonon interaction of the last term in Eq.\ (\ref{HC}).
It is worth mentioning that a similar impurity-mediated
electron-phonon interaction has been used to
account for the leading $T^2$ correction
to residue resistivity of simple metals~\cite{sInelasticImpurity,sMahan}.
Here, the direct scattering of electrons by phonons is neglected, which
was found to open new tunneling channels, and yield even,
positive correction to the conductance~\cite{sCaroli}. It does not account for
the ZBA. The electron tunneling Hamiltonian is given by
\begin{eqnarray}
H_{\mbox{\tiny T}}=\sum\limits_{kk's}\left(T_{kk's}^{\mbox{\tiny D}}a^{\mbox{\tiny L}\dagger}_{ks}
a^{\mbox{\tiny R}}_{k's}+\mbox{h.c.}\right)+\sum\limits_{lks}\left(
T_{lks}^{\mbox{\tiny IL}}a^{\mbox{\tiny L}\dagger}_{ks}d_{ls}+\mbox{h.c.}\right)
+\sum\limits_{lks}\left(
T_{lks}^{\mbox{\tiny IR}}a^{\mbox{\tiny R}\dagger}_{ks}d_{ls}+\mbox{h.c.}\right)\ ,
\label{HT}
\end{eqnarray}
where the first term represents direct tunneling of electrons between the electrodes,
and the second and third terms represent transfer of electrons between the electrodes
and the impurities.

To treat the electron-phonon interaction in a nonperturbative manner,
the Lang-Firsov transformation $e^{{\cal S}}$ with
${\cal S}=-\sum_{q\lambda ls}(M_{q\lambda}/\omega_{q\lambda})
d_{ls}^\dagger d_{ls}e^{i{\bf q}\cdot
{\bf R}_l}(b_{q\lambda}-b_{-q\lambda}^\dagger)$~\cite{sLangFirsov} is performed to
diagonalize the electron-phonon interaction.
The transformed Hamiltonian $e^{{\cal S}}He^{-{\cal S}}$ is the same as
the original one, except that the last
term in $H_{\mbox{\tiny C}}$ vanishes, and operators $d_{ls}$
and $d_{ls}^\dagger$ in $H_{\mbox{\tiny T}}$
are replaced with $d_{ls}X_{l}$ and $d^\dagger_{ls}X^{\dagger}_{l}$,
respectively, with $X_l=\exp[\sum_{q\lambda}(M_{q\lambda}/\omega_{q\lambda})
e^{i{\bf q}\cdot{\bf R}_l}(b_{q\lambda}-b_{-q\lambda}^\dagger)]$.
In this nonperturbative treatment, we see clearly that
electron tunneling via the impurities causes phonon emission
or absorption, which reversely renormalizes the electron tunneling matrix
elements via the impurities.

The path-integral
formulism on the closed-time path~\cite{sFeynman,sLYChen} is
employed to study this nonequilibrium problem. By integrating
out the electron variables in the two electrodes,
the retarded (advanced) Green's function
for electrons on impurity $l$ is obtained in frequency domain as
$\hat{G}^{\mbox{\tiny C}}_{l,r(a)}(\omega)=[\omega - E_0 + J
\hat{\mbox{\boldmath{$\sigma$}}}\cdot{\bf S}_l
-\hat{\Sigma}_{l,r(a)}(\omega)]^{-1}$.
Here, $2\times 2$ matrix representation in electron
spin space has been employed. The retarded self-energy
$\hat{\Sigma}_{l,r}(\omega)=\hat{\Sigma}^{\mbox{\tiny L}}_{l,r}(\omega)
+\hat{\Sigma}^{\mbox{\tiny R}}_{l,r}(\omega)$ is a diagonal matrix with diagonal
elements given by
\begin{eqnarray}
\Sigma_{ls,r}^{\mbox{\tiny L(R)}}(\omega)=\frac{1}{2}
\rho_{s}^{\mbox{\tiny L(R)}}\vert T^{\mbox{\tiny IL(R)}}_{l}\vert^2
\int d\omega'
\Bigl\{[1-f_{\mbox{\tiny L(R)}}(\omega + \omega')]B_>(-\omega')
+f_{\mbox{\tiny L(R)}}(\omega + \omega')B_{>}(\omega')\Bigr\}\ ,\label{SELF1}
\end{eqnarray}
and the advanced self-energy
$\hat{\Sigma}_{l,a}(\omega)=[\hat{\Sigma}_{l,r}(\omega)]^\dagger$.
Here, as usual, the momentum and spin dependence in the transmission
matrix element $T_{lks}^{\mbox{\tiny IL(R)}}$ is neglected~\cite{s6}.
$\rho_{s}^{\mbox{\tiny L(R)}}$ and $f_{\mbox{\tiny L(R)}}(\omega)$
are the electron DOS and Fermi distribution function
of the left (right) electrode, respectively. $B_>(\omega)$ is the Fourier transform of
$-i\langle X_{l}(t)X_{l}^\dagger(0)\rangle$, whose general expression
can be found in reference~\cite{sMahan}.

The tunneling current $I$ is calculated from the rate of the charges
flowing out of the emitting electrode.
The total conductance $G(V)=I/V$
is obtained as $G(V)=G^{\mbox{\tiny D}} + G^{\mbox{\tiny I}}(V)$,
where $G^{\mbox{\tiny D}}=
2\pi e^2\vert T^{\mbox{\tiny D}}\vert^2
\mbox{Tr}(\hat{\rho}^{\mbox{\tiny L}}\hat{\rho}^{\mbox{\tiny R}})$
is the direct tunneling conductance independent of bias voltage, and
\begin{eqnarray}
G^{\mbox{\tiny I}}(V)&=&
-\frac{N_{\mbox{\tiny I}}e}{2\pi V}
\int d\epsilon\int d\omega_1\int d\omega_2
B_{>}(\omega_1)B_{>}(\omega_2)
\Bigl\{f_{\mbox{\tiny L}}(\epsilon+\omega_1)
\left[1-f_{\mbox{\tiny R}}(\epsilon-\omega_2)\right]\nonumber\\
&-&f_{\mbox{\tiny R}}(\epsilon+\omega_1)
\left[1-f_{\mbox{\tiny L}}(\epsilon-\omega_2)\right]\Bigr\}
\overline{\vert T^{\mbox{\tiny IL}}_{l}
T^{\mbox{\tiny IR}}_l\vert^2
\mbox{Tr}\left[\hat{\rho}^{\mbox{\tiny L}}
\hat{G}^{\mbox{\tiny C}}_{l,r}(\epsilon)
\hat{\rho}^{\mbox{\tiny R}}
\hat{G}^{\mbox{\tiny C}}_{l,a}(\epsilon)
\right]}\ ,\label{CondJ}
\end{eqnarray}
comes from the tunneling via $N_{\mbox{\tiny I}}$ impurities.
The bias dependence of the tunneling conductance is embodied in
$G^{\mbox{\tiny I}}(V)$.
In Eq.\ (\ref{CondJ}), the last factor
is averaged over the positions and spin orientations of the impurities,
$\hat{\rho}^{\mbox{\tiny L}}=
[(\rho_\uparrow +\rho_\downarrow)+(\rho_\uparrow -\rho_\downarrow)\hat{\sigma}_z]/2$,
and $\hat{\rho}^{\mbox{\tiny R}}=
[(\rho_\uparrow +\rho_\downarrow)\pm(\rho_\uparrow -\rho_\downarrow)\hat{\sigma}_z]/2$
with sign $+$ for parallel alignment and $-$ for antiparallel
alignment. The two electrodes are considered of the same material
with electron DOS $\rho_\uparrow$ and $\rho_\downarrow$
for majority and minority spins, respectively.

There are two distinct resonant energies $E_{-}=E_0-JS$ and $E_{+}=E_0+JS$,
at which the tunneling conductance given in Eq.\ (\ref{CondJ}) will be
substantially enhanced.
In this work, we confine ourselves to the off-resonance case, which
is more probable in reality. The energy differences
$\vert E_{\mbox{\tiny F}}-E_{\pm}\vert$ with $E_{\mbox{\tiny F}}$ the
Fermi level are considered to be much greater than the broadenings of the resonant states.
Besides, the bias voltage $V$ is also
relatively small $\vert eV\vert < \vert E_{\mbox{\tiny F}}-E_{\pm}\vert$.
In this case, $\hat{\Sigma}_{l,r(a)}(\epsilon)$ in
$\hat{G}^{\mbox{\tiny C}}_{l,r(a)}(\epsilon)$ can be neglected.
The last factor in Eq.\ (\ref{CondJ}) is then divided into two parts:
$\overline{\vert T^{\mbox{\tiny IL}}_{l}
T^{\mbox{\tiny IR}}_l\vert^2}$ and $\overline{\mbox{Tr}[\hat{\rho}^{\mbox{\tiny L}}
\hat{G}^{\mbox{\tiny C}}_{l,r}(\epsilon)
\hat{\rho}^{\mbox{\tiny R}}
\hat{G}^{\mbox{\tiny C}}_{l,a}(\epsilon)
]}$. The former is a constant after averaged
over the impurity position ${\bf R}_l$,
and the latter involves only the average over
the orientation of ${\bf S}_l$. The energy of the magnetic field used to
produce the parallel alignment of the FM electrodes, usually
several ten or hundred oersteds ($g_{\mbox{\tiny L}}\mu_{\mbox{\tiny B}}SB\lesssim 0.01$meV),
is much smaller than the entropy gain due to
randomizing an impurity spin even
at experimentally low temperatures, e.g., 1K [$k_{\mbox{\tiny B}}T\ln(2S+1)\sim 0.1$meV].
Therefore, it is assumed that the impurity spins are randomly oriented
for both parallel and antiparallel alignments. Moreover, one can verify that
the results will not change significantly, if the magnetic field is strong
enough to align all the spins ferromagnetically in the parallel alignment.

The function $B_>(\omega)$ represents the spectrum density
of phonon emission and absorption. For simplicity, three-dimensional
Debye's model is used to describe the acoustic phonons.
A dimensionless electron-phonon coupling constant $g$
is defined as $g=3\sum_{q\lambda}
\vert M_{q\lambda}\vert^2\omega_{q\lambda}/\omega_{\mbox{\tiny D}}^3$
with $\omega_{\mbox{\tiny D}}$ the Debye frequency.
Thus, at zero temperature we have
\begin{equation}
iB_{>}(\omega)=\int dt\exp\left\{i\omega t
- g\left[\mbox{Cin}(\omega_{\mbox{\tiny D}}t)
+i\mbox{Si}(\omega_{\mbox{\tiny D}}t)\right]\right\}\ ,\label{B_GREATER}
\end{equation}
where $\mbox{Cin}(x)=\int_0^x dt[1-\cos(t)]/t$ and
$\mbox{Si}(x)=\int_0^x dt\sin(t)/t$ are the cosine and sine integral
functions~\cite{sEuler}.
$G^{\mbox{\tiny I}}(V)$ via impurities calculated
by use of Eqs.\ (4) and (\ref{B_GREATER})
is plotted in Fig.\ 1 for nonmagnetic tunnel junctions.
Since $\lim_{g\rightarrow 0}[iB_>(\omega)]=2\pi\delta(\omega)$,
$\lim_{g\rightarrow 0}G^{\mbox{\tiny I}}(V)=G_{0}^{\mbox{\tiny I}}$ is
a constant independent of $V$, which is actually the maximum
of $G^{\mbox{\tiny I}}(V)$. From Fig.\ 1,
it is clear that $G^{\mbox{\tiny I}}(V)$ vanishes at $V=0$
for any nonzero $g$, indicating
that tunneling via impurities is totally suppressed,
and $G^{\mbox{\tiny I}}(V)$ increases as the voltage $\vert V\vert$
is elevated. The stronger is the electron-phonon coupling $g$, the slower
the conductance $G^{\mbox{\tiny I}}(V)$ increases with $\vert V\vert$.
\begin{figure}
\includegraphics{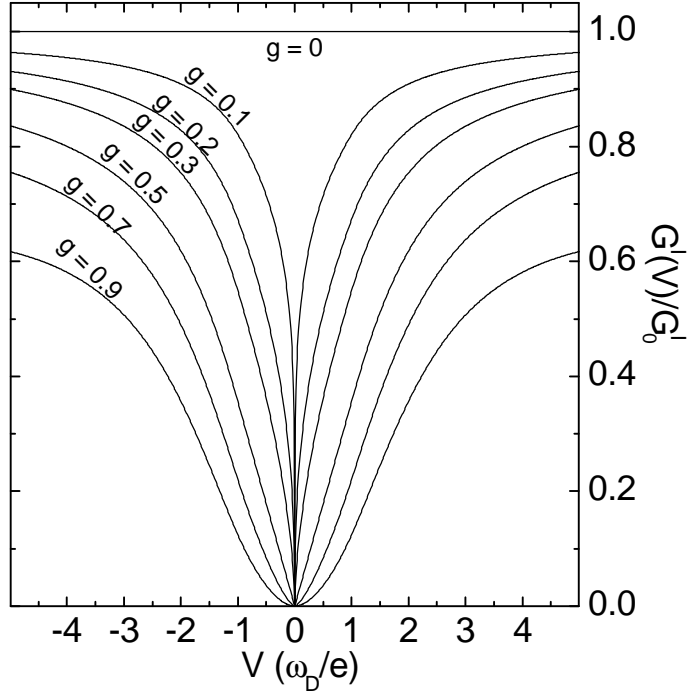}
\caption{ Normalized impurity conductance of nonmagnetic tunnel
junctions as a function of bias voltage at zero temperature. Here
$J=0$ and $\rho_{\uparrow}=\rho_{\downarrow}$. }
\end{figure}

To get some insight into low bias behavior of the
tunneling conductance, we derive the low-$\omega$
approximation of Eq.\ (\ref{B_GREATER})
\begin{equation}
i\omega_{\mbox{\tiny D}}B_{>}(\omega)\simeq \alpha\theta(\omega)
(\omega/\omega_{\mbox{\tiny D}})^{-(1-g)}\ ,\label{B_GREATER_APPX}
\end{equation}
where $\alpha = 2\Gamma(1-g)\sin(g\pi)e^{-g\gamma}$ with $\gamma=0.5772$
the Euler constant, and $\Gamma(z)=\int_0^\infty dtt^{z-1}e^{-t}$
is the ordinary gamma function~\cite{sEuler}. The unit step function
$\theta(\omega)$ indicates that at zero temperature only
emission of phonons is possible.
Substituting Eq.\ (\ref{B_GREATER_APPX})
into Eq.\ (\ref{CondJ}), we obtain an interesting result
$G^{\mbox{\tiny I}}(V)\sim \vert V\vert^{2g}$. For $g\le 0.5$ ($g>0.5$),
its slope $dG^{\mbox{\tiny I}}(V)/dV\sim \vert V\vert^{2g}/V$
is discontinuous (continuous) at $V=0$.

In Fig.\ 2, the resistances for parallel alignment ($R_{\mbox{\tiny P}}$)
and antiparallel alignment ($R_{\mbox{\tiny AP}}$) as well as
$\mbox{TMR}=(R_{\mbox{\tiny AP}}-R_{\mbox{\tiny P}})/R_{\mbox{\tiny AP}}\ $
of magnetic tunnel junctions are shown as functions
of the bias voltage. It is found that $R_{\mbox{\tiny AP}}$ always
decreases faster than $R_{\mbox{\tiny P}}$, resulting in
a decline of TMR. This behavior can be understood by the following
argument. At $V\simeq 0$, the direct tunneling dominates. For antiparallel
alignment, the majority-spin (minority-spin) band in the left electrode is
only connected to the minority-spin (majority-spin) band in the right
electrode, leading to a relatively high resistance. With increasing
$\vert V\vert$, the impurity tunneling increases.
Electron hopping
via the magnetic impurities causes a mixing of the two spin channels,
which effectively connects up the majority-spin bands in the two electrodes.
As a consequence, the resistance drops rapidly.
For $g=0.4$, the TMR is discontinuous in slope at $V=0$.
According to Fig.\ 2, either increasing the electron-phonon coupling $g$ or
decreasing relatively the tunneling via impurities weakens the ZBA.
In this theory, the energy scale of the ZBA is from zero to some $\omega_{\mbox{\tiny D}}$.
In view of that $\omega_{\mbox{\tiny D}}$ in most materials ranges from 10meV to 100meV,
the characteristic energies of the ZBA observed in tunnel junctions with
Ge~\cite{s1}, NiO~\cite{s10} and AlAs~\cite{add3} as insulating barriers are
consistent with our theory. For FM/Al$_2$O$_3$/FM, $V_{1/2}$ is greater than
100meV~\cite{s5,s6,s7,s8,s9}. The impurity density
might be low in these later junctions so that the ZBA is not prominent.
According to a simplified model for impurity tunneling~\cite{sShengNew},
we estimate that the impurity effect and
the ZBA will become observable as the impurity number per unit cross section reaches
$1/\mu m^2$ or higher.
\begin{figure}
\includegraphics{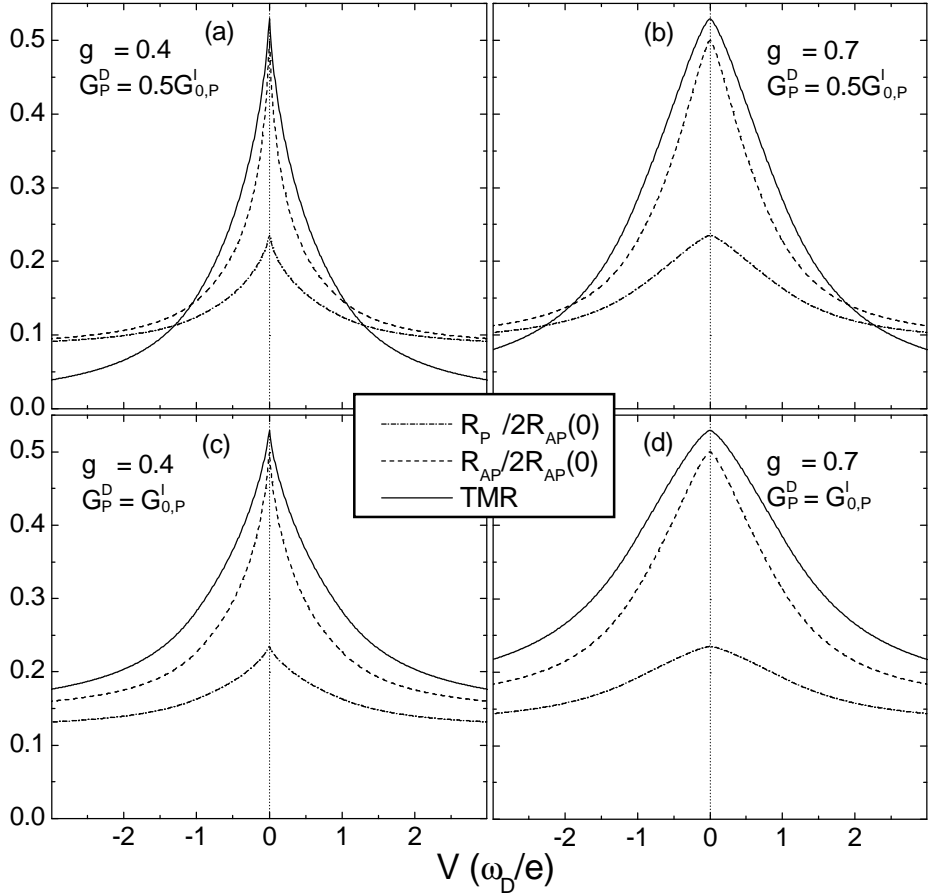}
\caption{Calculated results for $R_{\mbox{\tiny P}}$,
$R_{\mbox{\tiny AP}}$, and TMR as functions of bias voltage for
different $g$. The resistances are normalized by $2R_{\mbox{\tiny
AP}}(V=0)$. Here, $E_0=E_{\mbox{\tiny F}}$, and spin polarization
$(\rho_\uparrow-\rho_\downarrow)/(\rho_\uparrow+\rho_\downarrow)=0.6$.
}
\end{figure}

The present theory as well as the earlier theories~\cite{s6,add1,add2,add4,add11}
all predict that $G(V)$ for the tunnel junctions has a power-law
dependence on small bias voltage, i.e., $G(V)-G(0)\sim \vert V\vert^p$.
The low-temperature
curves of the conductance (resistance) and TMR exhibit
clear cusp-like feature at zero bias in
most experiments~\cite{s5,s6,s7,add7,add3}, and
non-cusp-like feature in a few experiments~\cite{add4}.
Mathematically, the cusp-like (non-cusp-like)
feature requires that the slope of the
conductance $dG(V)/dV\sim \vert V\vert^p/V$ be discontinuous (continuous)
at $V=0$, and so $p\le 1$ ($p> 1$). As mentioned in the introduction,
all the earlier theories~\cite{add1,add2,add4,add11} predicted constant
$p>1$, except for the theory of
Zhang $et$ $al.$~\cite{s6}, where $p\equiv 1$.
In the present theory, $p=2g$ may
be either smaller or greater than unity,
depending on the strength of the impurity-mediated
electron-phonon interaction, which is more
flexible for understanding both the cusp-like
and non-cusp-like features within a unified theoretical framework.
Our theory might be further verified directly by measurements of isotope
effect. Let us assume that the insulating barrier is
made of certain oxide. With gradual replacement
of the normal $O^{16}$ atoms with isotope $O^{18}$ atoms,
the mass $M$ in a lattice cell
increases, and the phonon frequency and Debye frequency
change as $\omega_{q\lambda}\sim M^{-1/2}$ and
$\omega_{\mbox{\tiny D}}\sim M^{-1/2}$. We notice
$\vert M_{q\lambda}\vert^2\sim (M_{\mbox{\tiny I}}
\omega_{q\lambda})^{-1}\sim M^{1/2}$, where $M_{\mbox{\tiny I}}$
is the impurity mass rather than the cell mass~\cite{sInelasticImpurity,sMahan}.
Consequently, the conductance exponent $2g$ in the present
theory will increase continuously
as $2g\sim M^{3/2}$ with the oxygen isotope enrichment.
In contrast, the constant conductance exponents
predicted by the earlier theories are expected
not to change in this case.

D. Y. X thanks the support from the National Natural Science Foundation of
China under Grant No. 10374046. D. N. S was supported by ACS-PRF \# 36965-AC5,
Research Corporation Grant CC5643, and the NSF grants DMR-00116566
and DMR-0307170.

\end{document}